\documentclass[12pt]{article}
\usepackage{amsfonts,amsbsy}
\newcommand{\be}{\begin{equation}}
\newcommand{\ee}{\end{equation}}
\newcommand{\bea}{\begin{eqnarray}}
\newcommand{\eea}{\end{eqnarray}}

\newcommand{\NC}{N}
\newcommand{\NSS}{{\cal N}}
\newcommand{\SUN}{SU(\NC)}

\newcommand{\ZN}{\mathbf{Z}_{\NC}}

\newcommand{\Fge}{\mathbf{\lambda}}
\newcommand{\Tr}{\mbox{Tr} \, }
\newcommand{\Haar}{\mbox{d}\mu_{\scriptscriptstyle H}}
\newcommand{\Per}[1]{\mathbf{S}_{#1}}
\newcommand{\Real}{\mathbb{R}}
\newcommand{\cC}{\mathcal{C}}
\newcommand{\FuM}{U}
\newcommand{\AdM}{\mathcal{U}}
\newcommand{\ADM}{\mathcal{U}_{\ell}}
\newcommand{\idper}{\mathsf{e}}
\newcommand{\lpper}{\mathsf{L}_+}
\newcommand{\Trg}[2]{\mathcal{T}(#2)^{#1}}

\newcommand{\cL}{{\cal L}}
\newcommand{\cI}{{\cal I}}
\newcommand{\cE}{{\cal E}}

\newcommand{\Latt}{\Lambda}
\newcommand{\GF}{{\cal Z}}
\newcommand{\Ge}{g_{\ell}}
\newcommand{\pab}{\tau}
\newcommand{\pver}{\vartheta}
\newcommand{\pcont}{\rho}

\input{epsf.tex}

\begin{document}

\vskip -4cm

\begin{flushright}
FTUAM-99-5

IFT-UAM/CSIC-99-6
\end{flushright}

\vskip -0.3cm

{\Large
  \centerline{{\bf Large $\NC$ corrections to the strong coupling  }     }
\centerline{{\bf behaviour of $\SUN/\ZN$ lattice gauge theories }}
\vskip 0.3cm

\centerline{ 
  A. Gonz\'alez-Arroyo{$^{\dag}$}\footnote{{\em e-mail:} tony@martin.ft.uam.es},  
 and C. Pena{$^{\ddag}$}\footnote{{\em e-mail:} carlos@martin.ft.uam.es},  
  }}
\vskip 0.3cm

\centerline{{$^{\dag}$}{$^{\ddag}$}Departamento de F\'{\i}sica Te\'orica C-XI and}
\centerline{{$^{\dag}$}Instituto de  F\'{\i}sica Te\'orica C-XVI}
\centerline{Universidad Aut\'onoma de Madrid,}
\centerline{Cantoblanco, Madrid 28049, SPAIN.}
\vskip 10pt

\vskip 0.5cm

\begin{center}
{\bf ABSTRACT}
\end{center}
We derive a formula for the large $\NC$ behaviour of the expectation values
of an arbitrary product of 
Wilson loops in the adjoint representation. We show the consequences 
of our formula  for the study of the large $\NC$  strong coupling behaviour of 
$\SUN/\ZN$ pure gauge theories, and theories with matter fields in 
the adjoint representation. This allows us to calculate large $\NC$ corrections
to the gluino condensate and meson propagators in the lattice version of 
Supersymmetric Yang-Mills. Applications to the strong coupling behaviour of
the Kazakov-Migdal model are also given.
\vskip 0.5 cm
\begin{flushleft}

\end{flushleft}

\newpage

\section{Introduction}
$\SUN$ group integration formulas at large $\NC$ have attracted interest since 
the late 70's. They occur in the study of large $\NC$  lattice gauge theories
and in connection with matrix models. The latter have found a wide range of 
applications over the years in areas going from 2-dimensional quantum gravity
to 4-dimensional Yang-Mills theory. The class of  matrix models involving 
large $\NC$ group integrations are  the  unitary matrix 
models (see Ref.~\cite{camporossi} for a recent review). 

Within lattice gauge theories,  $\SUN$ group integrations arise 
naturally when studying strong coupling expansions of lattice gauge theory
 models, which simplify in the large $\NC$ limit. 
The most standard method~\cite{effaction} uses an effective action approach
 centered around the one-link integral. Our motivation for this paper 
arose in the context of $\NSS=1$ SUSY Yang-Mills on the lattice, an interesting 
testing ground for non-perturbative methods, and  for which some
 Monte Carlo data are available~\cite{montvay}. The model  contains
lattice gauge fields and quarks (gluinos) in the adjoint representation of the
group. When studying the model at strong coupling one notices that   
 the effective potential method
cannot be carried through, lacking an equivalent one-link integral formula 
in the adjoint representation.  Our approach~\cite{SC} was based on the
 hopping parameter 
expansion expressed as a sum of random walks on the lattice~\cite{hoppingrandom,combi}. 
Then, one has to perform the gauge field  integration  of an expression 
containing products of traces of Wilson loops in the adjoint representation. 
It turns out, as will be shown in this paper, that the leading large $\NC$
contribution is rather trivial from the point of view of large $\NC$ 
integration. It arises from loops which are pure-spikes (or pure backtrackers)
and that actually do not depend on the gauge fields. 

Another instance  in which group integrals of quantities in the adjoint 
representation have acquired relevance is in the context  of 
{\em induced QCD}~\cite{kazakovmigdal}. The Kazakov-Migdal model has 
scalar quarks in the adjoint representation coupled to a $\beta=0$ 
lattice gauge field. A wide variety of techniques have been applied to this model. 
The strong coupling (large mass) region solution of the gaussian model was obtained by Gross~\cite{gross}.
Again, as realised later by Khoklachev and Makeenko~\cite{khomak}, the role of the large $\NC$ integration
over the gauge variables is simply to restrict the sum over random walks to 
pure-spike paths. Their contribution does not depend on the gauge fields,
 so that it can be taken outside the integral.

In this paper we will address the problem of computing the leading 
non-trivial large $\NC$ integral of a product of traces of Wilson loops 
in the adjoint representation. The result is  more easily presented in the 
form of the evaluation of a generating function (Eq.~(\ref{GenFun})) for 
adjoint Wilson loops. The main result is presented in Eq.~(\ref{ExpGenFun}). 
The formula is obtained by expanding the exponential, performing the integration 
and resumming the result, so that it is only valid for sufficiently small
lattice couplings $s(\gamma)$, i.e. it is the strong coupling expression of the 
generating functional. The result is the exponential of a quantity 
which is of order $1$ ($\sim \NC^0$). This contrasts with the typical $\NC^2$ behaviour 
for the fundamental representation. Indeed, in applications to 
strong coupling expansions of theories with matter fields,  
there is an additional   $\NC^2-1$ term arising from 
trivial Wilson loops which are proportional to the trace of the unit matrix in 
this representation. This term comes from the pure-spike paths mentioned 
previously. Thus, the terms we are calculating here give large $\NC$ subleading 
contributions to the observables of these theories. We emphasise nevertheless
 that our calculation gives the leading term in which the actual details of 
the group integration play any role.

The paper is organised as follows. The main quantity we are calculating 
is given in Eq.~(\ref{GenFun}) of Section 2, and the result for it is given in 
Eq.~(\ref{ExpGenFun}). The rest of the section contains the outline of the derivation, 
with details and terminology collected in the two appendices
(this includes a useful diagrammatic formalism to deal with the computation of
group invariants). 
Readers not interested in the derivation  might jump directly to Section 3,
where the formula in question is put to work in calculating large $\NC$
corrections to the strong coupling values of observables in $\SUN/\ZN$ 
lattice gauge theories, with or without matter fields. Explicit formulas are
given for the lattice version of $\NSS=1$ Supersymmetric Yang-Mills theory. 
This includes the large $\NC$ corrections to the gluino condensate and the leading non-OZI
 contributions to the  scalar meson propagator. Similar calculations can be carried on for 
the gaussian Kazakov-Migdal model. As an example, we show how from our formulas one can deduce the
leading $\NC$ behaviour of the propagators of the $p$-scalar bound states. The masses of these 
states are explicitly given.
 Finally, the paper is ended with a short conclusion section.

\section{Group integration formulas}

We consider an $\SUN$ gauge theory defined on a hypercubic, $d$-dimensional
infinite lattice $\Latt$. To any link  $\ell$ of the lattice  there is associated a
group  element $\Ge$. Most expressions that we will use  involve the 
matrices corresponding to the adjoint representation of the group $\AdM_A(\Ge)\equiv \ADM$.
 Regarding geometrical objects on the lattice, 
such as lattice paths, we will adopt  the terminology and conventions of Ref.~\cite{combi}.
Given a closed lattice path $\gamma$,  $\Tr(W_A(\gamma))$  labels  the trace of the 
corresponding Wilson loop  evaluated in the adjoint representation.  

Our main goal in this section is to  compute,   to leading order in $\NC$, the result of 
integrating over the group variables $\Ge$ an arbitrary   product of traces of adjoint Wilson
 loops. 
This result is best collected in the form of the 
 following generating function:
\be
\label{GenFun}
\GF(s) = \prod_{\ell}\left( \int \Haar(\Ge) \right) \exp \{ \sum_{\widetilde{\gamma}}
s(\widetilde{\gamma})\, \Tr(W_A(\widetilde{\gamma})) \} \ \ .
\ee
In this expression, $\Haar(\Ge)$ is the Haar measure on the group variable $\Ge$, and  the product
extends to  all the  links $\ell$ of the lattice.
The sum in the exponent runs over all equivalence classes of closed paths. Each equivalence class
 $\widetilde{\gamma}$ contains all paths such that the traces of the adjoint Wilson loops computed 
along them coincide for an arbitrary gauge field configuration. The common value of the trace 
of the adjoint Wilson loop is labelled $\Tr(W_A(\widetilde{\gamma}))$. The arguments of the 
generating function are the real parameters $s(\widetilde{\gamma})$. Notice that $\GF(s)$
can also be looked at as the partition function of a certain kind of pure gauge theory. The parameters 
$s(\widetilde{\gamma})$ would in that case appear as the (lattice) couplings of the
corresponding action. Although the action 
is not the most general $\SUN/\ZN$ invariant action, other terms would give superfluous information. 
From this point of view  the result of this section can be interpreted as the evaluation of the 
partition function of this 
pure gauge model in the strong coupling region (small  values of $s(\widetilde{\gamma})$).

The expansion of the exponential in $\GF$ gives us the integrals of traces 
that we are interested in:
\be
\GF(s)=\sum_{n=0}^{\infty}\frac{1}{n!}\sum_{\widetilde{\gamma}_1} \ldots \sum_{\widetilde{\gamma}_n}
s(\widetilde{\gamma}_1) \ldots s(\widetilde{\gamma}_n) \
\cE(\widetilde{\gamma}_1 \ldots \widetilde{\gamma}_n) \ \ ,
\label{GFExp}
\ee
with 
\be
\label{main}
\cE(\widetilde{\gamma}_1 \ldots \widetilde{\gamma}_n) =
\prod_{\ell} ( \int \Haar(\Ge))\, \Tr(W_A(\widetilde{\gamma}_1)) \ldots \Tr(W_A(\widetilde{\gamma}_n)) \ \ .
\ee
To compute the previous expression one must first select a representative of the class $\widetilde{\gamma}$.
 We will choose a  closed path without
 spikes (backtracking parts). It is always possible to reduce an arbitrary path
 by eliminating its
spikes in a systematic way and  reach a path which is called {\em simple}
in Ref.~\cite{combi}. As we will see later there is more than one simple 
path within each class, but anyone will do. Let us, from now on, use the same 
symbol $\widetilde{\gamma}$  to refer to the class and to its  
representative simple path.  
In the rest of this section we will show the main steps in the  evaluation
of  expression~(\ref{main}). Many of the technical details are collected in 
Appendices A and B.   

A possible approach would be to express the adjoint Wilson loops in terms of 
fundamental representation ones and then use the formulas available for the 
latter. Instead we would use a procedure which could allow the calculation of 
traces of adjoint Wilson loops and other gauge invariant objects systematically to all
orders in $1/\NC$.
What we do first is to express the integrand in Eq.~(\ref{main}) as a sum of 
products of the elements of the adjoint representation matrices $\ADM$:
\be
\label{main2} 
\cE(\widetilde{\gamma}_1 \ldots \widetilde{\gamma}_n) = 
 \sum_{a_i,b_i}^{} \delta(\pcont)^{a_1 \ldots a_L b_1 \ldots b_L} \prod_{\ell} \left( \int \Haar(\Ge) \right)
\AdM_{\ell_1}^{a_1 b_1} \ldots \AdM_{\ell_L}^{a_L b_L} \, \, .
\ee
The sum extends over the $2 L (\NC^2-1)$ possible values of the indices. For simplicity we will 
refer to the indices jointly by the  set $\cI \equiv \{a_1,b_1, \ldots a_L,b_L\}$. The integer 
$L$ gives the length  of the total joint path 
$\widetilde{\gamma}_1 \ldots \widetilde{\gamma}_n$. The labels $\ell_1 \ldots \ell_L$ 
refer to the links of this joint path taken in an arbitrary order and where a given link can 
occur more than once in the sequence. Finally,  the symbol 
  $\delta(\pcont)$ contains the contraction of the indices necessary to produce the 
  traces given in Eq.~(\ref{main}). It is given by a  product of $L$ Kronecker deltas.
It is clear that which index is contracted with which depends on the naming of the 
indices and on the geometry of the paths. This information can be coded into an application 
$\pcont:\cI \rightarrow \cI$ which maps the contracted indices onto each other 
($\pcont(c)=c'\, ; \pcont(c')=c$). Since the application is bijective it can be seen as an element 
of the group $\Per{2L}$ of permutations of $2L$ indices, having  the properties of a {\em pairing}.
It is clear from the definition of the index set $\cI$, that there is another pairing
 $\pab \in \Per{2L}$ naturally defined on our  system.
It  is the  application  which maps the two indices of a given link matrix onto each other:
$\pab(a_i)=b_i\,;\ \pab(b_i)=a_i$.

The next step in the calculation is the evaluation of the group integrals involved in 
Eq.~(\ref{main2}).  Our procedure  will be to 
express the adjoint representation matrices  in terms of the fundamental
representation ones, and then to make  use of the available formulas for
that  case. Here we will just explain and make use of the results, collecting the details
of the derivation within Appendix~A.
First, one can look at  the result of integrating  over a single  link variable. 
The result can be expressed  as a linear combination of invariant group tensors,
with coefficients depending on $\NC$. These tensors have indices within those elements of $\cI$ 
which belong to the matrices corresponding to this  link. The result of integrating over all link
variables can be expressed similarly as a linear combination of invariant tensors, this time involving 
all the indices of the set $\cI$. Thus we can write:
\be
\label{interm}
\cE(\widetilde{\gamma}_1 \ldots \widetilde{\gamma}_n) = \sum_{c_i} 
\delta(\pcont)^{c_1 \ldots c_{2L}} \sum_{\pver \in {\cal S}_{\Gamma}} C_{\NC}(\pver)\, 
\Trg{c_1 \ldots c_{2L}}{\pver} \ \ ,
\ee
where $\Trg{c_1 \ldots c_{2L}}{\pver}$ is an invariant tensor and  $C_{\NC}(\pver)$ the 
corresponding  $\NC$-dependent coefficient. The indices $c_i$ simply label the elements 
of $\cI$. The symbol $\pver$ stands for a permutation ($\pver: \cI \rightarrow \cI$). 
In Appendix~A (Eq.~\ref{trazas}) we describe how one can associate to any permutation of a certain 
type an invariant tensor ($\pver \longrightarrow {\mathcal{T}}(\pver)$). There, one can also see
 that if the permutation has a nonvanishing number of 1-cycles, the corresponding tensor vanishes.
 It is clear 
by the construction that both the invariant tensor and its coefficient 
factorise into a  product of contributions of  single links. This and other properties
(like the absence of 1-cycles) restrict the space of possible $\pver$. The full set labelled 
 ${\cal S}_{\Gamma}$ defines a certain subgroup of $\Per{2L}$. 
We will leave out its  characterisation for later and  continue with the outline of the 
procedure.  

The final step consists in performing the summation over the values of the indices, 
which is equivalent to contracting
the colour indices of the invariant tensor with the $\delta(\pcont)$.  Defining
\be
\label{defdiag}
D_{\NC}(\pver,\pcont) \equiv \sum_{c_i} \delta(\pcont)^{c_1 \ldots c_{2L}}\  \Trg{c_1 \ldots c_{2L}}{\pver}
\ee
we finally get:
\be
\label{maincompact}
\cE(\widetilde{\gamma}_1 \ldots \widetilde{\gamma}_n) = \sum_{\pver \in {\cal S}_{\Gamma}} C_{\NC}(\pver) D_{\NC}(\pver,\pcont) \ \ .
\ee
The result is  expressed in terms of the coefficients $C_{\NC}(\pver)$ and $D_{\NC}(\pver,\pcont)$. Explicit 
forms for these  quantities  can be 
traced back from Eqs.~(\ref{defjco}) and (\ref{defjtr}) in Appendix~A.
The main point now is to analyse the behaviour of these coefficients as $\NC 
\rightarrow \infty$.

 The large-$\NC$ behaviour of the coefficients $C_{\NC}(\pver)$ follows
from the result for the corresponding coefficient in the one-link integral,
Eq.~(\ref{cnnlargeN}); one gets:
\be  
\label{jcolargeN}
C_{\NC}(\pver) = A(\pver)\ \left( \frac{2}{\NC} \right)^L + \mbox{ subleading terms} \ \ .
\ee
where the leading prefactor  $A(\pver)$ is zero for all but a restricted class of 
permutations $\pver$.  It is not hard to characterise this subset within ${\cal S}_{\Gamma}$. It is given by 
those permutations $\pver$ satisfying: $\pver \pab \pver \pab = \idper$, where $\idper$ is the identity mapping.    
For these permutations the coefficient $A(\pver)$ is equal to $1$. This result rests on the 
work of Ref.~\cite{barsgreen}. 

The machinery to deal with the contractions of invariant tensors entering the definition of $D_{\NC}$,
and extract their leading $\NC$ behaviour, is set up in
Appendix~B, whither we refer for details. The method follows by setting up a Feynman-diagram interpretation of
 each term in the sum Eq.~(\ref{maincompact}). For that we need to enter into the structure of the 
invariant tensors ${\mathcal{T}}(\pver)$. These are products of traces of products of the group generators in the  
fundamental representation (cf. Eq.~(\ref{blobdec})). To any $p$-cycle contained in the permutation
 $\pver$ there corresponds
an invariant tensor of the form  $\Tr(\Fge^{c_1} \ldots \Fge^{c_p})$ (where $\Fge^{c_i}$ is  
a group generator). It is clear why the tensor vanishes if there are $1$-cycles. 
We look at these traces as vertices of a Feynman diagram where the indices play the 
role of the legs. The vertices are not symmetric with respect to the exchange of the legs 
but only cyclic-symmetric. This form of vertices is characteristic of planar diagrams in large 
hermitian matrix models, for example.  
Thus, for the full $\pver$ one obtains a bunch of vertices with a total
of $2L$ legs. The contractions encoded in $\pcont$ can be seen as propagators joining pairs of 
legs of different or the same vertices. Thus to every term in  Eq.~(\ref{maincompact}) one can associate 
a (vacuum) Feynman diagram with   $V$
vertices and $L$ lines (note that $L$, which is the total number of pairs of colour indices
in Eq.~(\ref{main2}), is the only topological quantity of the diagrams being equal for all the
$\pver$ and $\pcont$).
We further call $K$ the number of connected components of  the diagram.

In Appendix~B we prove that the large $\NC$ maximal leading  contribution of a diagram with $K$ connected 
components, $V$ vertices, and $L$ lines is given by $m\, 2^{-L}\NC^{2K-V+L}$, where $m$ is an integer
that can be zero or one. Corrections go like inverse powers of $\NC^2$. If we combine 
 this with~(\ref{jcolargeN}), we obtain that the leading term coming from
  $ C_{\NC}(\pver) D_{\NC}(\pver,\pcont)$  goes at most as $\NC^{2K-V}$.

Now, from the topological constraints it is clear that $1 \le K \le V$. Therefore, the maximum
value of $2K-V$ would occur when $K=V$. This means that all propagators go from one vertex to 
itself. To understand the implications of this we need to make use  of some of the properties of the vertices appearing in such a diagram and 
its relation to the original set of lattice paths. Since vertices arise from link integration they can only 
contain indices pertaining to matrices of the same link. Furthermore, vertices only mix the first indices of the 
matrices among themselves and the second among themselves (see Appendix A for details). Hence, a contraction 
between two indices of the same vertex comes from a contraction of the link matrix with itself. But since the first index
is contracted with the first or the second with the second,  we have either $\AdM_{\ell}^t \AdM_{\ell}$ or 
 $\AdM_{\ell} \AdM_{\ell}^t$, which due to the orthogonal nature of the adjoint representation matrix allows 
to eliminate the two matrices altogether. In terms of the original paths a self contraction corresponds  to the path 
moving in one direction forward and backward consecutively. This gives what we call spikes. But remember that our 
original paths were simple, and hence had  no spikes. In conclusion, the restriction to simple
paths tells us that there are no self-contractions of our diagrammatic vertices, corresponding to the situation 
when these vertices come from normal ordered operators.

We then  conclude that all connected components of the diagram have at least two vertices. Hence,
the leading term arises from diagrams for which $2K=V$ and which give a contribution
of order one as $\NC \rightarrow \infty$. Finally, we arrive at
\be
\label{mainfinal}
\cE(\widetilde{\gamma}_1, \ldots\widetilde{\gamma}_{n})=\nu_0 + {\cal O}(\frac{1}{\NC^2}) \ \ ,
\ee
where $\nu_0$ is an integer summing the contribution from all the leading 
order diagrams.  This is the main result that was needed in Ref.~\cite{SC}
to justify the restriction to {\em pure-spike} paths. In what follows we will proceed to compute the 
value of $\nu_0$.

First of all let us study what can we learn about the graphs giving  nonvanishing 
leading contribution (${\cal O}(\NC^0)$). We have not yet given the condition that has to be 
satisfied by $\pver$ to give a nonvanishing leading contribution to $D_{\NC}(\pver,\pcont)$.
This occurs (see Appendix B)  when the permutation $\pver$ satisfies the relation:
$\pver \pcont \pver \pcont = \idper$.  This is very similar to the characterisation coming from the 
link integration,  with the replacement of $\pab$ by $\pcont$.  Our final number $\nu_0$ 
comes from counting the number of possible 
permutations $\pver$ within ${\cal S}_{\Gamma}$ satisfying both relations simultaneously.
Furthermore, the set ${\cal S}_{\Gamma}$ is  characterised by saying that it contains all permutations 
having no $1$-cycles and such that all cycles contain indices belonging to matrices of the 
same link and to either the first or the second index of these matrices  (but not both). 

To understand the implications  of the previous conditions, let us return to the connection between the 
original paths $\widetilde{\gamma}_i$ and the permutation $\pcont$. Any of the paths is an ordered 
sequence of matrices. Since  the matrices have two indices we get a classification of indices into 
two sets: those appearing as a first or a second index of the matrices. Notice that if the link is followed in the 
reverse direction the definition of what index is first or last is exchanged. Any path is then associated 
to two equal length closed sequences of indices. It corresponds to two of the cycles of the permutation 
$\pcont \pab$. The two cycles can be connected by means of $\pab$ (or $\pcont$). Now, from the conditions 
characterising the leading $\pver$, we conclude: 
\be
\pcont \pab = \pver \pcont \pab \pver^{-1}\quad .
\ee
This means that    $\pver$ must map complete  cycles of $\pcont \pab $ into equal length 
cycles. Given the identification of these cycles with the original paths, we see that 
a leading $\pver$ just maps  complete paths of equal length among themselves. But since $\pver$
only connects indices belonging to the same lattice point, we conclude that the actual paths 
connected by $\pver$ have to coincide. There are a few very simple conclusions one can extract from 
this result. The first is that there is a factorisation property satisfied by  the integral
 $\cE(\widetilde{\gamma}_1, \ldots\widetilde{\gamma}_{n})$. Its value factorises into a 
product of contributions of distint paths. This follows from a similar factorisation property for 
 $\pver$, since indices from different paths cannot be connected by it, despite the fact that the 
paths might overlap partially. Thus, the final  number of maximal permutations $\nu_0$ factorises 
as well. One might have guessed this factorisation from the standard arguments in the large 
$\NC$ limit; notice, however, that the factorisation is not complete: as we will see, the integral 
of a given Wilson loop to the power $n$ is not the $n$-th power of the integral of a single Wilson
loop, so that the result is not classical.

It only remains, therefore, to compute $\nu_0$ for the case of a single Wilson loop $\Tr(W_A(\widetilde{\gamma}))$
taken to the power $n$.
To do this, it is convenient to express the path $\widetilde{\gamma}$ in the form $(\gamma')^{
\omega}$,
where $\omega$ is a positive integer. This notation means that $\widetilde{\gamma}$
is built up by describing $\gamma'$ sequentially $\omega$ times; $\gamma'$ itself must not be
reducible equivalently. We will refer to $\omega$ as the {\em winding number} of the path,
and write it as $\omega(\widetilde{\gamma})$. These considerations are necessary, as our
results will depend explicitly on $\omega(\widetilde{\gamma})$.

For our path configuration we have that $\pcont \pab$ has $n$ cycles  of length 
$l(\widetilde{\gamma})$ (the length of the path). The permutation $\pver$ maps 
one cycle into other preserving the ordering within each cycle. Thus, it is specified
by giving the corresponding mapping of the indices at a given space point. Naively 
one would say that the number is then $n!$. In fact this has to be corrected by two facts: 
the first is that we have to eliminate all permutations $\pver$ having $1$-cycles;
the second is that  one must take into account that if $\omega(\widetilde{\gamma})$ is larger than $1$, the mapping 
has to respect the cyclic ordering in which $\widetilde{\gamma}$ describes these points. 
This last issue can be  dealt with  by arranging the indices belonging to the same 
path into cycles of length $\omega(\widetilde{\gamma})$. In a compact mathematical form
we have that, given  a permutation $\tau'$ having $n$ $\omega$-cycles,  we want to count
how many permutations $\pver'$ without 1 cycles of the $n\omega$ indices satisfy: 
$\pver' \tau' \pver'^{-1} =\tau'$. Let us call this number  $P^{n}_{\omega}$. 
The action of $\pver'$ is simply the mathematical statement of a change of names of the indices, 
so that without the requirement of no $1$-cycles, the solution would be simply $n! \omega^n$.
This comes because we must map complete cycles into complete cycles, so we have $n!$ possibilities. 
Then after having selected the mapping of the cycles, there are $\omega$ possible assignements 
for each cycle (it is enough to specify the mapping of 1 element of the cycle and the rest 
would follow): this gives the $\omega^n$.
Subtracting the permutations which have $1$-cycles takes a little more effort but presents 
no real problem.   
 As a matter of fact, it is easier and more compact to determine the expression for the 
generating function first and deduce from it the expression of the quantity $P^{n}_{\omega}$.
For the generating function corresponding to a single path one obtains:
\be
\label{GenFunOne}
 \prod_{\ell}\left( \int \Haar(\Ge) \right) \exp \{ 
s(\widetilde{\gamma})\, \Tr(W_A(\widetilde{\gamma})) \} = 
\frac{\exp\{-s(\widetilde{\gamma})\} }{(1-\omega(\widetilde{\gamma}) s(\widetilde{\gamma}))}\quad ,
\ee  
For completeness we give an expression for the numbers $P^{n}_{\omega}$ themselves:
\be
P^{n}_{\omega} = n! \sum_{p=0}^{n} \frac{(-1)^p}{p!}\, \omega^{n-p} \quad .
\ee

Finally, we combine the result Eq.~(\ref{GenFunOne}) for all the paths via the factorisation 
property and arrive at the desired  leading contribution to the generating
functional Eq.~(\ref{GenFun}):
\be
\label{ExpGenFun}
\GF(s) = \exp\left( - \sum_{\widetilde{\gamma}} \left(  s 
(\widetilde{\gamma}) + 
\log(1-\omega(\widetilde{\gamma})s(\widetilde{\gamma}))\right) \right)
\ee
which is the main result of this section. As explained in the introduction 
the exponent is of order $\NC^0$. Had we added a contribution to the initial action 
 proportional to the trace of the unit matrix in the adjoint representation, the 
term would have remained unchanged by the integration (being a constant)  and would have given a contribution 
to the exponent of order $\NC^2-1$  (the dimension of the adjoint representation).
Although trivial, this remark is important in practice because such a term actually 
arises in many applications. It can be associated to the trivial path of zero length.

\section{Application to strong coupling expansions}
In this section we will use the group integration 
formula obtained in the last section to derive results which are relevant for 
the behaviour of $\SUN/\ZN$  gauge theories on the lattice, with or without 
matter fields, at strong coupling and large $\NC$. These include pure gauge theories,
gauge theories with fermions in the adjoint representation, a particularly 
interesting case being the lattice version of $\NSS=1$ Supersymmetric Yang-Mills~\cite{montvay}, and 
also gauge-Higgs models with scalars in the adjoint representation like the 
Kazakov-Migdal model~\cite{kazakovmigdal}. In all cases, after integrating out the matter fields via 
a hopping parameter expansion, one ends up with an effective action of the type 
given in Eq.~(\ref{GenFun}). Our integration formula Eq.~(\ref{ExpGenFun}) allows us to write 
immediately the expression for the free energy in the strong coupling region at large $\NC$.
In what follows we will give the explicit formulas in some of the cases that 
have been studied previously in the literature.

Consider first the case of $\SUN/\ZN$ pure gauge lattice theories. For this case the
 consequence of our formulas can be easily deduced. As mentioned at the beginning of 
Section 2, the generating function $\GF(s)$ can be seen as the partition function of a pure gauge model 
where the couplings $\beta_A(\widetilde{\gamma})$ are identified with $s(\widetilde{\gamma})$. 
The free energy of the model is then given by the exponent of the right hand side of Eq.~(\ref{ExpGenFun}). 
As a particular example we can write down the case of the Wilson-type plaquette action model. 
The free energy per unit volume of the model is given by:
\be
\label{freepure}
F(\beta_A)/Volume = \frac{d(d-1)}{2} \left( - \beta_A - \log(1- \beta_A) \right) \ \ ,
\ee
which has a singularity at $\beta_A=1$. Actually, the position of the singularity 
coincides with the position of the third-order phase transition point in the 
2-dimensional lattice gauge theory~\cite{samuel}. In higher dimensions 
the model is known to have a first order phase transition line at $\beta_A<1$~\cite{mak2,samuel}.  
Thus, the strong coupling phase depicted by Eq.~(\ref{freepure}) extends into 
the metastable region as is usually the case  for these transitions. There is one atypical behaviour
of the free energy Eq.~(\ref{freepure}) which explains why this formula was not found before. 
This is the fact that it is of order $1=O(\NC^0)$. Typically the free energy goes like $\NC^2$, 
as for example is the case in this same model above the first order phase transition point. 
The different $\NC$ dependence of the 
free energy above and below the transition suggests that the position of the transition is given  
by the vanishing of  the weak coupling free energy. This is indeed the case in 2 dimensions. 
An  $\NC^2$ behaviour is also shown by the free energy of the mixed fundamental-adjoint action for 
any non-zero value of the fundamental coupling~\cite{samuel}. Thus,  in older work 
the free energy  (\ref{freepure}) of this model  is taken to be zero~\cite{brihayerossi}. 
Remarkably, from Eq.~(\ref{ExpGenFun}) one can easily read out the free energy 
 for an arbitrary $\SUN/\ZN$ invariant action.

Let  us now consider the addition of matter fields in the adjoint representation. The
general features and behaviour is similar for scalar or fermionic matter. For definiteness 
we will illustrate the formulas restricting to the fermionic case. Indeed, our 
concern with $\NSS=1$ SUSY  Yang-Mills triggered our interest in the problem. 
Here we will follow the procedure of Ref.~\cite{SC}. Performing  the integration over the quark                     
(gluino) fields leads to a determinant. In the large mass (small hopping) region, this determinant 
 can be written as the
exponential of an effective action expressable as a sum over closed paths on
the lattice:
\be
\label{effaction} 
S_{eff}=\eta_f \displaystyle{\sum_{x \in \Lambda} \sum_{l=1}^{\infty} \sum_{\gamma \in {\cal S}_l(x)}}
\frac{1}{l}\Tr(W(\gamma))\Tr(\Gamma(\gamma)) \ \ ,
\ee
where ${\cal S}_l(x)$ is the set of closed paths of length $l$ with origin in  the lattice point $x$,
and $\Gamma(\gamma)$ is a spin matrix associated to the path, whose explicit form we will
detail later (cf. Eq.~(\ref{mspin})). The factor $\eta_f$ takes into account
the different possibilities for the nature of the fermion field: $\eta_{Majorana}=1/2$,
while $\eta_{Dirac}=1$.

In Ref.~\cite{combi} it is shown  how it is convenient and possible to 
rearrange the summation over paths into a sum over classes of paths 
labelled by what is 
called a {\em simple path}.  A simple path is a closed contour that has no 
spikes or backtracking pieces. It is easy to see that for all paths associated 
with the same simple path the trace of the Wilson loop is exactly the same,
 so that they  actually belong to a  single  equivalence class of the type
defined at the beginning of Section 2. However, the classes considered in this paper
contain more than one simple path and its corresponding classmates. 
The reason is that in  Ref.~\cite{combi} closed 
paths are defined as ordered sets of links, which have therefore a  first and 
a last element.  It is clear, however, that which link is taken as the first one is 
irrelevant  for the gauge field dependence, being  given by a trace. This is 
just a technical point, but essential to get the numbers right.
To compensate for this point one has simply to multiply by a counting factor 
which gives the number of links that can be chosen as first element. 
Curiously this factor is not the length of the path $l(\widetilde{\gamma})$ but 
 $l(\widetilde{\gamma})/\omega(\widetilde{\gamma})$, where $\omega(\widetilde{\gamma})$ 
is the winding number factor defined in the last 
section.  When the path goes round a given loop $\omega(\widetilde{\gamma})$ times, it is 
clear that the relevant factor is the number of links in the basic loop, and 
not in the total path.  In addition, the orientation of the path is 
irrelevant for the gauge field dependence, so that  our  present classes contain each path and 
its reverse.  In summary, we stress that the classes of paths defined in Section~2 contain a 
basic simple path, its reverse and all the $l(\widetilde{\gamma})/\omega(\widetilde{\gamma})$
simple paths obtained by selecting a point along the path (as origin).

Finally, one can rewrite the effective action (Eq.~(\ref{effaction})) in the form of the 
 exponent of Eq.~(\ref{GenFun}),  
 with coefficients $s(\widetilde{\gamma})$ which depend on the basic 
parameters of the original theory. These  coefficients are ${\cal O}(1)$ in $\NC$.
One can include in them the contribution of the pure gauge action.
Restricting for definiteness to one flavour  adjoint fermions at zero gauge couplings,
which contains the lattice version of $\NSS =1 $ supersymmetric Yang-Mills, one 
obtains the following   coefficients:
\bea
\label{sspin}
  s(\widetilde{\gamma}) &=& \frac{\tau(\widetilde{\gamma})}{\omega(\widetilde{\gamma})} \quad ,\\
\label{deftau}
\tau(\widetilde{\gamma}) &\equiv& \frac{\eta_f}{(1-\xi)^{l(\widetilde{\gamma})}} \left( \Tr(\Gamma(\widetilde{\gamma}))+ \Tr(\Gamma(\widetilde{\gamma}^{-1})\right) \quad ,
\eea 
where $\gamma^{-1}$ is the reverse of the path $\gamma$, and the matrix $\Gamma(\widetilde{\gamma})$
is the spin matrix of the path, given 
by the product along the path of the link matrices:
\be
\label{mspin}
\kappa\ (r \mathbf{I}_s - \gamma_{\alpha})\quad .
\ee
In the previous expression  $\kappa$ is the hopping parameter, $r$ the Wilson parameter, and 
the spin matrices $ \gamma_{\alpha}$ are just the Dirac matrices up to a sign, positive  if
 the path traverses the link with positive orientation and negative  if not. The factor 
$1-\xi$ in Eq.~(\ref{deftau}) is the result of resumming all paths within each class, with $\xi$ given by:
\be
\label{xidef}
\xi=\frac{1-\sqrt{1-4(2d-1)\kappa^2(r^2-1)}}{2} \ \ ,
\ee
where $d$ is the space-time dimension.

With the previous expression and formula~(\ref{ExpGenFun}), one can obtain the free energy of the 
model at large $\NC$:
{\setlength \arraycolsep{2pt}
\bea
\nonumber
F&=&\eta_f\, (N^2-1) \ F_0 \  Volume - \\
&&\sum_{\widetilde{\gamma}} \Bigg( \frac{\tau(\widetilde{\gamma})}{\omega(\widetilde{\gamma})}+
\log( 1- \tau(\widetilde{\gamma})) \Bigg) \ \ ,
\label{FreeEn}
\eea
}
where the sum runs over the equivalence classes of paths $\widetilde{\gamma}$ defined in Section~2.

We see that our result provides the ${\cal O}(1)$ contribution to the free energy. This is not however
the leading piece, which is given by the term proportional to $F_0$, and is of order $\NC^2$.
The latter was given in Ref.~\cite{combi} and its value is:
\be
F_0=\eta_f\, \bigg( -d \log(1-\xi) + (d-1) \log(1-\frac{2d}{2d-1}\xi) \bigg) \Tr(\mathbf{I}_s) \ \ ,
\ee
where  $\mathbf{I}_s$ is the identity matrix in spin space.
This term arises from the paths whose gauge field contribution is proportional to the trace of the unit matrix. This 
occurs for the so-called pure-spike paths, which are paths such that after eliminating all 
backtracking parts we are left with a single point (a path of zero length). This contribution is, 
however, fairly trivial from the point of view of the group integration, so that the ${\cal O}(1)$ 
contribution is the first one in which the characteristics of group integration show up.

Obtaining a closed expression for the  ${\cal O}(1)$  contribution to the free energy 
 involves performing the sum over path classes explicitly. Similar resummations can be
 performed (see Ref.~\cite{combi} for a derivation of these 
resummations and for earlier references on the subject). The appearance of the factor $\omega(\widetilde{\gamma})$
in the formula has prevented us so far from obtaining such a closed expression.  

Other interesting quantities can be computed to the order in $\NC$ at which we are working. This includes 
the gluino condensate, which can be obtained by an appropriate differentiation of the free energy. 
As in the previous case, the leading term given in Ref.~\cite{SC} comes from pure-spike paths, 
 and is  (up to some multiplication factors) independent of the group representation in 
which the quarks are living. This is a particular manifestation of the triviality from the point of view of group 
integration of the leading term.  Finally, our expression is as follows:
{\setlength \arraycolsep{2pt}
\bea
\nonumber
\langle \bar{\Psi}(x)\Psi(x) \rangle &=& - \frac{(\NC^2-1) \Tr(\mathbf{I}_s)}{1-\frac{2d}{2d-1}\xi} - \\
&& \frac{1}{Volume} \sum_{\widetilde{\gamma}}  
 \frac{l(\widetilde{\gamma})\, \tau(\widetilde{\gamma})}{\eta_f\, (1-2\xi)}\left(-\frac{1}{\omega(\widetilde{\gamma})}+
\frac{1}{1-\tau(\widetilde{\gamma})}\right) \ ,
\label{ChiralCond}
\eea
}
where $l(\widetilde{\gamma})$ is the length  of any representative simple path.
The division by the volume in the previous expression is simply accounted for by picking a 
single simple path $\widetilde{\gamma}$ from the space of all space-time translates. 
The correction  is the first unquenched contribution. The leading 
one coincides with the result of   the quenched approximation~\cite{SC}.
To arrive at expression Eq.~(\ref{ChiralCond}) we have made use of the resummation formulas
of Ref.~\cite{combi}. The origin of the $1- 2\xi$ factor can be traced back to the 
rearrangement of sum over paths into sums over simple paths.

In a similar fashion, it is possible  to derive the  large $\NC$ correction to the meson and 
$p$-gluino propagators and spectrum, which are given in Ref.~\cite{SC}. We will not give the formulas explicitly since they are 
 notationally fairly lengthy. Instead we will provide the expression  of  a quantity 
 for which our present computation gives the leading 
large $\NC$ behaviour. This quantity  is  the non-OZI contribution to the meson propagator. 
In a theory with different equal mass flavours, this term  gives 
the main contribution to the mass difference between the non-singlet and singlet mesons.
It contains what in the continuum would be the contribution of the anomaly.
When expanding the fermion propagators in terms of paths, the non-OZI contributions come from 
2 closed paths originating at the 2 meson positions (see Ref.~\cite{SC} for details).
Due to the main factorisation property which 
we have obtained for Wilson loop integration at this order in $\NC$, the contribution is only non-zero 
when the two paths (classes) coincide. To the order in $\NC$ at which we are working, this term is non-zero 
both in the quenched approximation and in the full theory. For simplicity, we give the  contribution to the 
scalar propagator:
\bea
G_{non-OZI}(x,y) &=& \frac{1}{\eta_f^2\left(1- \frac{2 d}{2d-1}\xi \right)^2}
\sum_{\hat{\gamma}_1\hat{\gamma}_2  } 
\tau(\hat{\gamma}_1)\tau(\hat{\gamma}_2)\, \delta_{\widetilde{\gamma}_1  \widetilde{\gamma}_2}
\left( \frac{\omega(\widetilde{\gamma}) }{ 1- \tau(\widetilde{\gamma})} \right)^2
\label{propnonOZI}
\eea
where the sum now extends over all closed terse paths  $\hat{\gamma}_1$ with origin in $x$ and $\hat{\gamma}_2$
 with origin in $y$ (see Ref.~\cite{combi} for an explanation of the lattice path terminology and an account 
of the resummation formulas). The $\delta_{\widetilde{\gamma}_1  \widetilde{\gamma}_2}$ imposes that the two paths 
have the same common  
reduced simple path, which we label $\widetilde{\gamma}$. It is possible to eliminate the constraint 
given by the $\delta$ in the previous formula, but its treatment exceeds the purpose of this paper.

In a similar way one can study other models, like the Kazakov-Migdal model which has scalars in the 
adjoint representation. The gaussian model can be treated in a similar way as the fermionic model. 
In this model there are composites formed by $p$ scalars, which are the analog of the $p$-gluino 
bound states of the fermionic model~\cite{SC}. The operators which produce these states can be written in 
our notation as follows:
\be
\mathbf{O}_p(x)\equiv \phi_{a_1}(x) \cdots \phi_{a_p}(x)\, \Trg{a_1 \ldots a_p}{\lpper^{(p)}}\quad .
\ee
At strong coupling  the correlator of two operators at  different points (say $x$ and $y$) can be written as a sum 
over  $p$ paths ($\gamma_1 \ldots \gamma_p$) joining the fields  at the two different points. 
The gauge field dependence of such a contribution is as follows (we restrict  to correlators 
that are leading at large $\NC$):
  \be 
\label{gaugedep}
 \Trg{a_1 \ldots a_p}{\lpper^{(p)}} \ \Trg{b_p \ldots b_1}{\lpper^{(p)}} \,
(W_A(\gamma_1))^{a_1 b_1} \ldots  (W_A(\gamma_p))^{a_p b_p}\quad .
\ee
Now we have to evaluate its expectation value with respect to the gauge field in the large $\NC$ limit.
The quantity  is not  a product of traces of Wilson loops. To handle this in a simple way with our formulas,
we might add a new {\em big link} $\bar{l}$ going from $y$ to $x$ and its corresponding group element (as an
independent variable). Now, one can form 
$p$ closed paths by attaching to each of the original open paths $\gamma_i$ the big link $\bar{l}$ (note they all 
have $\omega=1$).
One can then compute the expectation value of the product of the corresponding Wilson loops. Clearly $s$ for these 
paths is zero, since none of the paths in the effective action contains the big link. The  large $\NC$ leading term of the 
connected expectation value is obtained for all paths belonging to the same class.  The corresponding 
contribution is the $p$-th derivative of the free energy in Eq.~(\ref{ExpGenFun}) at $s=0$, which is $(p-1)!$. 
What has this to do with the expectation value of Eq.~(\ref{gaugedep})? To see that, one has simply to integrate over 
the big link group variable. One gets  terms like Eq.~(\ref{gaugedep}) multiplied by $\left(\frac{2}{\NC}\right)^p$. 
Then one has to multiply by the number of different $p$-cycles, which is $(p-1)!$. Hence, finally, the 
expectation value of Eq.~(\ref{gaugedep}) to leading order is given by  $\left(\frac{\NC}{2}\right)^p$,
and is achieved when all the path classes $\widetilde{\gamma}_1 \ldots \widetilde{\gamma}_p$ coincide.
Similar results  can be obtained  in the fermionic model, thus justifying the treatment of $p$-gluinos of 
Ref.~\cite{SC}. As an application of this result one can proceed in a similar fashion as in Ref.~\cite{SC}
and deduce the values of the masses $M_p$ of the $p$-field bound states for this theory:
\be
\label{pgmasslarger}
\cosh(M_p) = \frac{(2d-1)^{p/2}}{2} \Bigg( \left(\frac{1-\xi}{\xi}\right)^{p/2}
+ \frac{1}{(2d-1)^{p-1}}\left(\frac{\xi}{1-\xi}\right)^{p/2} \Bigg) -\sigma \, ,
\ee
$\sigma$ being an integer number in $\{-(d-1), \ldots , (d-1)\}$, which accounts for
all the possible doubler modes (see~\cite{SC} for details), and $\xi$ is given by the same formula as before
(Eq.~\ref{xidef}) with $r^2 -1$ set to $1$.

\section{Conclusions}
In this paper we have studied the leading behaviour for large number of 
colours  of $\SUN$ group link integrals of products of traces of Wilson 
loops in the adjoint representation.  We have proven  that the
leading (non-trivial) result is ${\cal O}(1)={\cal O}(\NC^0)$. The corresponding 
coefficient has  been computed in Section~2. The result factorises 
into contributions of Wilson loops associated to inequivalent 
paths. This allows us  to compute up to this order and in the strong 
coupling (small $s(\gamma)$) region the corresponding 
generating function, given in Eq.~(\ref{ExpGenFun}) of Section~2. 

In Section~3
we have shown how this generating functional and its derivatives 
enter in expressions  which are relevant in the context of strong coupling
expansions for gauge theories with matter fields in the adjoint
representation. We have given the expression for the free energy of a pure gauge 
$\SUN/\ZN$ model at strong coupling. Adjoint  Wilson loop expectation 
values can be read out immediately for these theories. We then proceeded to show 
how the large mass expansion of theories with matter fields in the adjoint 
representation yields an effective action suitable for treatment with our 
formula. We illustrate the applications by calculating explicitly the large 
$\NC$ corrections to various quantities in the lattice version of Supersymmetric 
Yang-Mills. In all cases, to obtain a compact form, one is faced with a 
path resummation problem which at the moment we have been unable to solve. 
(The problem looks  tractable, though). We also showed that our present formulas 
can be used to prove the form of the leading large $\NC$ behaviour for 
$p$-gluino propagators and spectra given in Ref.~\cite{SC}. 
We then  took the case of the gaussian 
Kazakov-Migdal model to show how, similarly, one arrives at the large $\NC$ 
values of the masses of the $p$-particle bound states at strong coupling.

From the methodological point of view, our paper might also prove of interest 
in other contexts where the large $\NC$ limit is at play. Our  diagrammatical
 technique is non-standard and could help to complement 
other well-known techniques as loop equations, orthogonal polynomials, 
master field equation, etc. It allows the computation of  
 higher order corrections in $1/\NC^2$  keeping the Feynman 
diagram analogy. On the negative side, such a technique does not look
 appropriate for dealing with the weak coupling (large $s(\gamma)$) 
behaviour of the generating function.

\section*{Acknowledgements}
This work has been financed by the CICYT under grant AEN97-1678.
A.~G-A wants to thank the Theory Division at CERN for its 
hospitality during the last stages of this work. 
The authors acknowledge useful conversations with E. Gabrielli and J.
 Vermaseren.

\section*{Appendix A}

In this appendix we study group integrals with variables in the adjoint
representation of the gauge group. For this purpose, as explained in Section~2,
we will relate the matrices
in the adjoint representation to their fundamental counterparts, and afterwards
we will use known integration formulas for the latter. These will be first
applied to the computation of single link integrals, and then we will use the
result to evaluate the multilink integral in Eq.~(\ref{main2}), and reduce it
explicitly to the compact form in Eq.~(\ref{interm}).

Let us first establish some notations and conventions. We write $\Fge^a$ for the generators of
$\SUN$ in the fundamental representation. They close the group algebra:
\be
\label{algebra}
\lbrack \Fge^a,\Fge^b \rbrack = \imath f^{abc}\Fge^c \ \ ,
\ee
where the indices $a,b,c$ range from 1 to $\NC^2-1$. We choose the generators to be normalised as:
\be
\label{gennorm}
\Tr(\Fge^a \Fge^b) =\frac{1}{2} \, \delta^{ab} \ \ .
\ee
Furthermore, the completeness relation is written as:
\be
\label{completeness}
 (\Fge^a)^{ij} (\Fge^a)^{kl} = \frac{1}{2} \, \delta^{jk}\delta^{il} -
\frac{1}{2\NC} \, \delta^{ij}\delta^{kl} \ \ .
\ee

Now, the group integrals we want to compute have the general form:
\bea
\label{adjint}
\int \Haar(g) \, \AdM^{a_1 b_1} \ldots \AdM^{a_n b_n} \ \ ,
\eea
where $\AdM=\AdM_A(g)$ is the matrix corresponding to a given group element $g$ in the {\em adjoint} representation,
and $\Haar(g)$ is the Haar measure on $\SUN$.

The relation of $\AdM$ with the matrix $\FuM=\AdM_F(g)$ corresponding to the same group element in the fundamental
representation reads, with the conventions adopted above:
\be
\label{reladjfun}
\AdM^{ab}=2 \, \Tr(\FuM^{\dag} \Fge^a \FuM \Fge^b) \ \ .
\ee
This allows us to rewrite~(\ref{adjint}) as an integral for matrices
in the fundamental representation, whose indices are contracted with a
certain tensor built up of a product of generators. Before casting this
statement into an explicit equality, let us introduce the following
known formula for group integrals in the fundamental representation~\cite{barsgreen}:
\bea
&\int \Haar(g) \, \FuM^{i_1 j_1} \ldots \FuM^{i_n j_n} (\FuM^{\dag})^{l_1 m_1} \ldots (\FuM^{\dag})^{l_n m_n} = \nonumber \\
\label{fundint}
&~~~~~~~~~~~~~~\displaystyle{\sum_{\sigma,\sigma' \in \Per{n}}} \delta(j,\sigma(l)) \, 
\delta(m,\sigma'(i)) \, C_N^n(\sigma \circ \sigma') \ \ , \\
&\mbox{with}~~n<\NC \nonumber \ \ .
\eea
In this expression, $\Per{n}$ is
the permutation group of $n$ elements, and we have introduced the following notation:
\be
\label{defbigdelta}
\delta(j,\sigma(l)) \equiv \prod_{k=1}^{n} \delta_{j_k l_{\sigma(k)}} \ \ ,
\ee
where $l_{\sigma(k)}$ is the $k$-th element of $\sigma(\{l_1 , \ldots , l_n\})$,
with $\sigma \in \Per{n}$.

The coefficients
$C_N^n$ are class-functions on the space of permutations, i.e.,
$C_N^n$ is a map $\Per{n} \to \Real$ fulfilling
$C_N^n (\sigma \circ \sigma' \circ \sigma^{-1})=C_N^n (\sigma')$ for any
$(\sigma,\sigma')$ (the notation $\sigma \circ \sigma'$ stands for the composition
of two permutations), so that
the value $C_N^n (\sigma)$ only depends on the total number of cycles in the permutation
$\sigma$. For this purpose we call $\cC_k(\sigma)$ the number of $k$-cycles in $\sigma$,
and $\cC(\sigma) \equiv \sum_k \cC_k(\sigma)$. With this notation, it can be shown that the
following relation holds~\cite{barsgreen}:
\be
\label{coefcomp}
\delta_{\sigma' \tilde{\sigma}} = \sum_{\sigma \in \Per{n}} C_N^n(\sigma' \circ \sigma) \,
\NC ^{\cC(\sigma \circ \tilde{\sigma})} \ \ ,
\ee
which allows to determine $C_N^n(\sigma)$ recursively. Its general form turns out to be
$C_N^n(\sigma) = \frac{P(\NC)}{Q(\NC)}$, where $Q(\NC)=\NC^2(\NC^2-1) \ldots (\NC^2-(n-1)^2)$,
and $P(\NC)$ is a polynomial whose leading term is $\NC^{\cC(\sigma)}$. Hence, at large $\NC$ we get:
\be
\label{cnnlargeN}
C_N^n(\sigma) \stackrel{\NC \to \infty}{=} \frac{1}{\NC^n} \delta_{\sigma,\idper} + \mbox{ subleading terms} \ \ ,
\ee
where $\idper$ is the identity permutation, which maps a sequence to itself.

If we now merge Eqs.~(\ref{reladjfun}) and (\ref{fundint}), the result for~(\ref{adjint})
can be written as:
\be
\label{adjintsol}
\int \Haar \, \AdM^{a_1 b_1} \ldots \AdM^{a_n b_n} = 2^n \sum_{\sigma,\widetilde{\sigma} \in \Per{n}}
C_N^n(\sigma \circ \widetilde{\sigma}) \, \Trg{a_1 \ldots a_n}{\sigma} \, \Trg{b_1 \ldots b_n}{\widetilde{\sigma}} \ \ ,
\ee
where we have introduced the following notation:
\be
\label{trazas}
\Trg{a_1 \ldots a_n}{\sigma} \equiv
(\Fge^{a_1})^{j_1 k_1} \ldots (\Fge^{a_n})^{j_n k_n} \delta(k,\sigma(j)) = 
(\Fge^{a_1})^{j_1 j_{\sigma(1)}} \ldots (\Fge^{a_n})^{j_n j_{\sigma(n)}} \ \ ,
\ee
an implicit summation over matrix indices being understood. These objects are products of
traces of products of generators, and therefore invariant group tensors. To make this more
explicit, we can introduce the one-step-forward cyclic permutation of $m$ elements $\lpper^{(m)}$, defined
by $\lpper^{(m)}(a_i)=a_{i+1}$ for $i<m$ and $\lpper^{(m)}(a_m)=a_1$, such that:
\be
\label{trazabas}
\Trg{a_1 \ldots a_m}{\lpper^{(m)}} = \Tr(\Fge^{a_1} \ldots \Fge^{a_m}) \ \ .
\ee
Then we notice that each cycle in a generic permutation of $n$ elements $\sigma$, as
that appearing in~(\ref{trazas}), closes an $\lpper$ trace, so that we can write:
\be
\label{blobdec}
\Trg{a_1 \ldots a_n}{\sigma} = \prod_{i=1}^{\cC(\sigma)} 
\Trg{a_{\sigma_i(1)} \ldots a_{\sigma_i(\nu_i)}}{\lpper^{(\nu_i)}} \ \ ,
\ee
where $\sigma_i$ is a sub-permutation giving the index order inside the $i$-th cycle,
which is a $\nu_i$-cycle.

\vskip 0.5cm

Now we address the computation of Eq.~(\ref{main2}). If we substitute for each single-link
integration in it the result~(\ref{adjintsol}) we get:
{\setlength\arraycolsep{1pt}
\bea
\cE(\widetilde{\gamma}_1 \ldots \widetilde{\gamma}_n) &=&
2^L \sum_{a_i, b_i} \delta(\pcont)^{a_1 \ldots a_L b_1 \ldots b_L} \prod_{k=1}^{\cL} \Big[
\sum_{\sigma^{(k)},\widetilde{\sigma}^{(k)} \in \Per{n_k}} 
\nonumber \\
&&C_{\NC}^{n_k}(\sigma^{(k)} \circ \widetilde{\sigma}^{(k)})\, 
\Trg{a_1 a_2 \ldots a_{n_k}}{\sigma^{(k)}}\, 
\Trg{b_1 b_2 \ldots b_{n_k}}{\widetilde{\sigma}^{(k)}} 
\Big] \, \, ,
\label{expmain}
\eea
}
where the index $k$ labels the links which contribute at least one $\AdM$ in
Eq.~(\ref{main2}), the total number of these links being $\cL$
(note that the integral over a link carrying no $\AdM$ is 1). We call $n_k$ the
number of $\AdM$'s in~(\ref{main2}) lying on the link $k$; as we define in Section~2 the total number
of $\AdM$'s to be $L$, it is obvious that $\sum_{k=1}^{\cL} n_k = L$.

One can put the result in a more compact form by introducing  a permutation $\pver \in \Per{2L}$
combining the  set  $\lbrace \sigma^{(k)},\widetilde{\sigma}^{(k)}; k=1,\ldots,\cL \rbrace$.
It is important to notice that the set ${\cal S}_{\Gamma}$ of all the possible $\pver$
obtained in this way is, in general, only a subgroup of $\Per{2L}$, determined by the
structure of the joint path $\widetilde{\gamma}_1 \cup \ldots \cup \widetilde{\gamma}_n$.
Notice that the condition for $C_{\NC}(\pver)$
to be leading in $\NC$ is read from Eq.~(\ref{cnnlargeN}) to be
$\sigma^{(k)}=\widetilde{\sigma}^{(k)^{-1}}$ for any $k$; when written in terms of the
joint permutation $\pver$,
this condition is equivalent to $\pver \pab \pver \pab = \idper$, where $\pab$ is a pairing,
defined in Section 2, such that $\pab(a_i)=b_i,\pab(b_i)=a_i$.

Once $\pver$ is fixed, we can define:
\bea
C_{\NC}(\pver) &\equiv& 2^L \prod_{k=1}^{\cL} C_{\NC}^{n_k}(\sigma^{(k)} \circ \widetilde{\sigma}^{(k)})
\ \ ,
\label{defjco} \\
\Trg{c_1 \ldots c_{2L}}{\pver} &\equiv& \prod_{k=1}^{\cL}
\Trg{a_1 a_2 \ldots a_{n_k}}{\sigma^{(k)}} 
\Trg{b_1 b_2 \ldots b_{n_k}}{\widetilde{\sigma}^{(k)}} 
\ \ ,
\label{defjtr}
\eea
after renaming the colour indices as in Section~2.
It is these definitions, together with Eq.~(\ref{defdiag}), that give the final
form Eq.~(\ref{maincompact}).

\section*{Appendix B}

In this appendix we will study the large $\NC$ behaviour of scalars (invariants) formed by contraction
 of invariant tensors. This type of objects  appears in the definition 
of  the quantities $D_{\NC}$ entering in Eq.~(\ref{maincompact}). We will use the symbolic 
representation  for 
tensors as given  in~(\ref{trazas}).  For notational details we 
refer the reader  to  Appendix~A. Computational methods for calculating these quantities have been 
devised~\cite{vermaseren} due to their interest in different contexts. Our iterative technique is 
related  to the method of  Cvitanovi\'c~\cite{cvita}. 

First of all we will set up a Feynman diagram-like  representation  for the invariant tensors
$\Trg{a_1 \ldots a_m}{\sigma}$.  For that, one uses the decomposition of the permutation 
$\sigma$ into cycles (see Eq.~(\ref{blobdec})). The tensor corresponding to an $m$-cycle
$ \Trg{b_1 \ldots b_m }{\lpper^{(m)}}$ is the trace  of the product of $m$ group generators,
as shown in Eq.~(\ref{trazabas}).  Now, to each $\lpper^{(m)}$ tensor  we associate an $m$-legged vertex,
 where the external legs are attached to indices, oriented clockwise according 
to the orientation given by the cycle.  
In this way, a tensor associated to the permutation $\sigma$ is associated to  $\cC(\sigma)$
vertices. Furthermore it will prove useful to consider  vertices with no legs, associated to the
trace of the identity. This representation is illustrated by the following:

\hbox{
\hskip 3.5cm
\vbox{
$\displaystyle{\Trg{b_1 \ldots b_n}{\lpper^{(n)}}~~\longrightarrow~~~}$
\vskip 0.5cm
}
\hskip -9.5cm
\vbox{
\epsfxsize=2cm \epsffile{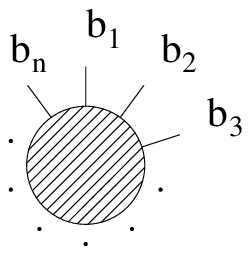}
}
}

\hbox{
\hskip 1cm
\vbox{
$\displaystyle{\Trg{b_1 \ldots b_n}{\sigma}~~~\longrightarrow~~~}$
\vskip 0.5cm
}
\hskip -10cm
\vbox{
\epsfxsize=7.33cm \epsffile{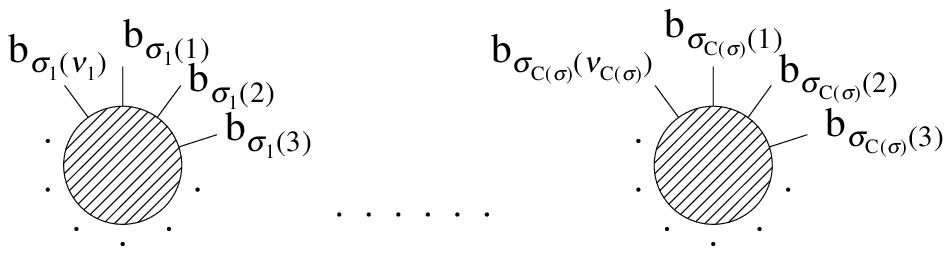}
}
}

\hbox{
\hskip 3.5cm
\vbox{
$\displaystyle{\Tr(\mathbb{I}) = \NC ~~~\longrightarrow~~~}$
\vskip 0.5cm
}
\hskip -9.5cm
\vbox{
\epsfxsize=2cm \epsffile{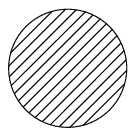}
}
}

The contraction of two indices of a tensor is represented in the diagrams by joining the
corresponding legs. Given a product of invariant traces with  contracted indices,
these prescriptions allow to represent  it by  a diagram consisting of a certain number
of vertices, joined by internal lines, and   possibly with some external lines representing unsaturated
indices of the tensor. For the scalar quantities we are interested in, the corresponding 
diagram is a vacuum diagram with no external legs. Notice, however, that the ordering of the legs in 
a vertex is relevant, up to cyclic permutations. As explained in Section~2 and used later, the actual 
pattern of contraction of indices in a tensor can be labelled by a {\em pairing} $\pcont$ of the indices, 
that is a permutation made up of only $2$-cycles.

Our  technique is based upon equations which express  the result of contracting   two indices
in an $(m+n+2)$-tensor as a linear combination of $(m+n)$-tensors. These equations can be obtained 
easily from the completeness relation of the group generators Eq.~(\ref{completeness}). We will refer 
to the basic equations as   {\em fusion} and {\em fission} rules, and they are given respectively by:
{\setlength\arraycolsep{1pt}
\bea
\sum_a \Trg{a b_1 \ldots b_m}{\lpper^{(m+1)}} \Trg{a c_1 \ldots c_n}{\lpper^{(n+1)}} &=& 
\frac{1}{2}\, \Trg{b_1 \ldots b_m c_1 \ldots c_n}{\lpper^{(m+n)}} - \nonumber \\
\label{fusion}
&&\frac{1}{2\NC} \, \Trg{b_1 \ldots b_m}{\lpper^{(m)}}\Trg{c_1 \ldots c_n}{\lpper^{(n)}} \\
\nonumber \\
\sum_a \Trg{a b_1 \ldots b_m a c_1 \ldots c_n}{\lpper^{(m+n+2)}} &=& 
\frac{1}{2} \, \Trg{b_1 \ldots b_m}{\lpper^{(m)}} \Trg{c_1 \ldots c_n}{\lpper^{(n)}} - \nonumber \\
\label{fission}
&&\frac{1}{2\NC} \, \Trg{b_1 \ldots b_m c_1 \ldots c_n}{\lpper^{(m+n)}} \ \ .
\eea
}

In our diagrammatic representation the fusion and fission rules can be depicted as:


\hbox{
\vbox{
\epsfxsize=4cm \epsffile{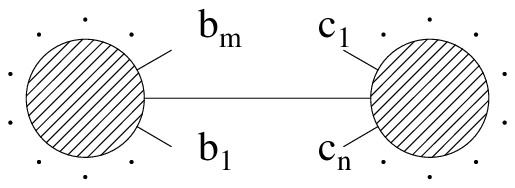}
\vskip -0.5cm
}
\hskip -0.7cm
\vbox{
$\displaystyle{=~~~\frac{1}{2}}$
}
\hskip -11.7cm
\vbox{
\epsfxsize=1.33cm \epsffile{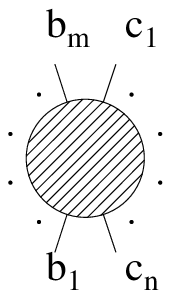}
\vskip -1cm
}
\hskip -0.5cm
\vbox{
$\displaystyle{-~~~\frac{1}{2\NC}}$
\vskip -0.2cm
}
\hskip -11.5cm
\vbox{
\epsfxsize=4cm \epsffile{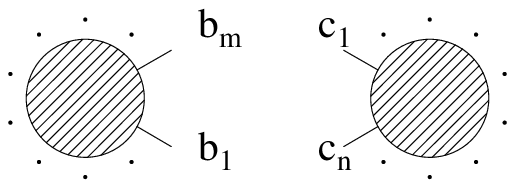}
\vskip -0.8cm
}
}
\vskip -1.5cm
\be
\label{diagfus}
\ee

\vskip 1cm

\hskip 0.5cm
\hbox{
\vbox{
\epsfxsize=2cm \epsffile{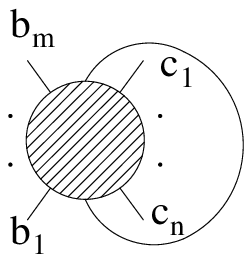}
\vskip -0.9cm
}
\hskip -0.5cm
\vbox{
$\displaystyle{=~~~\frac{1}{2}}$
}
\hskip -11.7cm
\vbox{
\epsfxsize=4cm \epsffile{blobx2.eps}
\vskip -0.9cm
}
\hskip -0.8cm
\vbox{
$\displaystyle{-~~~\frac{1}{2\NC}}$
\vskip -0.3cm
}
\hskip -11.5cm
\vbox{
\epsfxsize=1.33cm \epsffile{blobx1.eps}
\vskip -0.9cm
}
}
\vskip -1.5cm
\be
\label{diagfis}
\ee

\vskip 1cm

 It is easy to see that the repeated  application
of Eqs.~(\ref{diagfus}, \ref{diagfis}) allows the evaluation of all 
scalars formed by contracting the indices of an  arbitrary tensor. 
Our goal will be to extract information on the $\NC$ dependence from 
these equations. As an example and for later use, we give 
in Table~1 the value and associated diagram for all  independent tensor
contractions up to six indices. The corresponding diagrams have  $K=1$ 
connected components and   up to $L=3$ internal lines. 

If the number of indices of a tensor is odd, making all possible 
contractions would leave an unsaturated index. Applying the fusion-fission
rules would reduce it to a linear combination of tensors with one index. 
But the only one-index tensor $\Tr(\Fge^a)$ vanishes, and hence
the contribution of a diagram with a single external leg vanishes too. 
This allows to prove a result which will be needed in what follows:
the scalar corresponding to a one-particle reducible diagram vanishes;
by applying fusion and fission rules iteratively in each of those
two sides of the diagram connected by the reducible line we would reach 
 the form $B(\NC) \, \Tr(\Fge^a) \Tr(\Fge^a)$, which is zero q.e.d.

Now we will prove a result concerning the $\NC$ dependence of the scalars 
formed by contraction of a given tensor, expressed in terms of the topological
properties of the associated diagram. The result is that the leading 
$\NC$ behaviour of such a scalar is given by $m2^{-L}\NC^{2K-V+L}$, with
$m$ being either 0 or 1. It is easy to check this property directly for tensors 
with few indices. The tensor contractions  given in Table~1  and their 
multiplication allow  
to check it up to $L \le 3$. The  general proof will be done by induction,
making use of the fusion and fission rules.
Hence, we will start by assuming that the result has been proven for all scalars
obtained after the contraction of $L$ pairs of indices, and end up proving 
that the result is also true for $L+1$ contractions.

Consider then a scalar formed by performing $L+1$ contractions of the indices of a
$2L+2$ invariant tensor and its associated one-particle irreducible diagram. Let say it has
 $K$ connected components, $V$ vertices and  $L+1$ internal lines. Applying the fusion 
rule its contribution is given as a  linear combination  of the contribution of 
two diagrams with $L$ internal lines,   with coefficients $1/2$ and $-1/2\NC$.
The first diagram has $L$ internal lines, $V-1$ vertices and $K$ connected components;
thus, its leading term is of the form $m_1 2^{-L} \NC^{2K-V+L+1}$.
The second has $L$ internal lines, $V$ vertices and $K$ connected
components\footnote{It cannot have $K+1$ since this would contradict the hypothesis that the original diagram is 
one-particle irreducible.},
so that the leading contribution is $m_2 2^{-L} \NC^{2K-V+L-1}$, which is subleading with respect 
to the previous one. In summary, the behaviour of the initial scalar is given by 
 $m 2^{-(L+1)} \NC^{2K-V+(L+1)}$, as predicted by our result,
with $m=m_1$.

Now let us turn to the fission rule. As the original diagram has $K$ connected components,
$V$ vertices and $L+1$ internal lines, then the two subdiagrams in the decomposition
have $(K',V+1,L)$ and $(K,V,L)$
topological numbers respectively, with $K'$ being either $K$ or $K+1$. The leading
term of the original diagram is thus of the required form $m 2^{-(L+1)} \NC^{2K-V+(L+1)}$,
with $m=0$ if $K'=K$ and $m=m_1$ if $K'=K+1$. This completes the proof.

The previous result is used in Section~2, to prove that the coefficients $D_{\NC}(\pver,\pcont)$,
which are associated  to a  given  diagram, have the mentioned  large $\NC$ behaviour. This,
combined with the leading behaviour of the coefficients $C_N$, leads to the conclusion that 
the large $\NC$ dominant behaviour of the integrals $\cE$ (Eq.~(\ref{main})) can only come 
from diagrams made out of connected components with two vertices. From our previous results
we also know that the prefactor of the leading term is an integer counting the number of diagrams 
giving a non-vanishing maximal leading behaviour. For that purpose we need to complement our 
previous result by characterising those diagrams which have a maximal large $\NC$ behaviour ($m=1$).
Fortunately, we only need to do this for diagrams with two vertices $V=2$ (and no self-contractions),
making up one of the connected components of a possible maximal diagram.

\vspace{0.5cm}

In the remaining of  this Appendix, we will prove the statement made in Section~2,
namely that the two-vertex connected (and with no 
self-con\-trac\-tions) diagrams having maximal leading behaviour ($\NC^{L}$) are the ones satisfying 
the identity $ \pver \pcont \pver \pcont = \idper$, where $\pver$ is the permutation 
giving the vertex structure and $\pcont$ the permutation specifying the contractions. 
By the form of the diagram we know that $\pver$ is made of $2$ cycles of length $L$, 
and $\pcont$ of $L$ cycles of length $2$. Let us see what this means in diagrammatic 
terms. Take the legs of one  vertex and label one as $a_1$, then label the 
subsequent legs $a_{i+1}\equiv\pver(a_i)$. This means that the legs are ordered in a 
clockwise fashion. Now call $b_i\equiv \pcont(a_i)$ the indices of the other vertex (remember 
there are no self-contractions). The condition $ \pver \pcont \pver \pcont = \idper$
is equivalent to having  $\pver(b_i)=b_{i-1}$. This means that in the other vertex the ordering
of the indices is counter-clockwise. The corresponding diagram can then be drawn by joining the 
legs with  lines which do not cross: the diagram is planar.

 Using this graphical characterisation, it is easy to show that
a diagram coming from $\pver \pcont \pver \pcont \ne \idper$ can never have a leading term, of the form
studied above, with $m=1$. On the other hand, an exact formula can be derived for planar
diagrams by iterating the fusion rule; the result with $L \ge 2$ lines is:
\be
\label{V2planar}
\left(\frac{\NC^2-1}{2 \NC}\right)^L \left(1-\frac{1}{(1-\NC^2)^{L-1}}\right) \ \ ,
\ee
and obviously gives leading behaviour with $m=1$.

To prove that any nonplanar diagram is necessarily not leading, let us
depict a generic nonplanar diagram with $L=s+2$ lines in the following form
(which is always attainable),

\hbox{
\hskip 4.5cm
\vbox{
\epsfxsize=4cm \epsffile{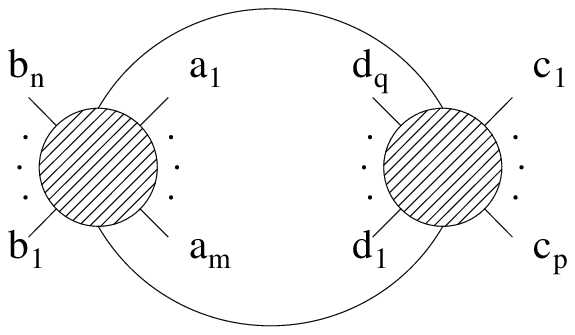}
\vskip 0cm
}
}

where the indexed legs are completely contracted between themselves (remember that no
self-contractions are allowed), and $s=m+n=p+q \ge 1$. The nonplanarity
condition enters as a constraint: both $a-c$ and $a-d$ contractions exist.

If a fusion is applied, and then a fission to the resulting first term in the rhs,
one gets:

\hbox{
\hskip 0.5cm
\vbox{
\epsfxsize=4cm \epsffile{nonp_lhs.eps}
\vskip -1.1cm
}
\hskip -0.7cm
\vbox{
$\displaystyle{= ~ \frac{1}{4}}$
}
\hskip -12.3cm
\vbox{
\epsfxsize=3.6cm \epsffile{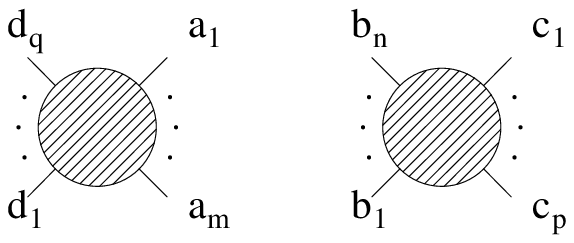}
\vskip -0.8cm
}
\hskip -0.5cm
\vbox{
$\displaystyle{ - ~~ \frac{1}{4\NC}}$
\vskip -0.2cm
}
\hskip -12cm
\vbox{
\epsfxsize=3.5cm \epsffile{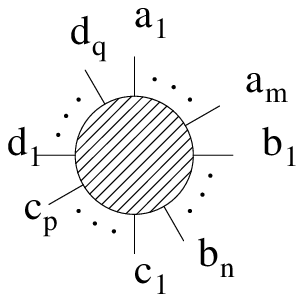}
\vskip -0.8cm
}
}
\vskip 0.5cm
\hbox{
\hskip 4.3cm
\vbox{
$\displaystyle{ - ~~ \frac{1}{2\NC}}$
\vskip 0.1cm
}
\hskip -12cm
\vbox{
\epsfxsize=3.5cm \epsffile{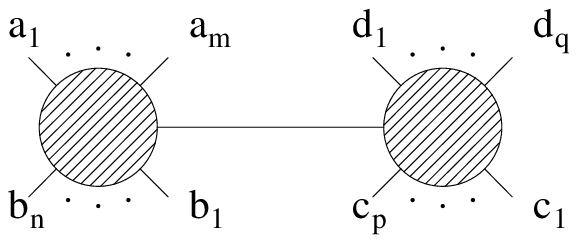}
\vskip -0.5cm
}
}
\vskip -1.5cm
\be
\label{nonplanar}
\ee

Now, the nonplanarity constraint prevents the first term in the rhs to have more
than one connected component. It is then immediate, after applying the rule for the
leading term to each diagram in the rhs, that the  leading behaviour for the nonplanar
diagram goes at most like  $ \NC^{L-2}$.

\newpage

\noindent Table 1: Explicit results for all topologically inequivalent,
nonvanishing, completely connected
diagrams corresponding to contractions of invariant traces with
up to 3 internal
lines are shown. Disconnected diagrams with $L \le 3$ can be built
as products of the ones shown here.

\begin{center}

\hbox{
\hskip 0.5cm
\vbox{
\epsfxsize=3.2cm \epsffile{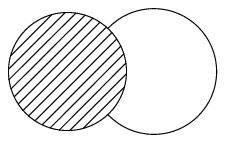}
\vskip -1.1cm
}
\hskip -5.2cm
\vbox{
$\displaystyle{=~~\Tr(\Fge^a \Fge^a)}$
}
\hskip -9.0cm
\vbox{
$\displaystyle{=~~\frac{\NC^2-1}{2}}$
}
}

\hbox{
\hskip 0.5cm
\vbox{
\epsfxsize=3.2cm \epsffile{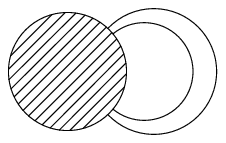}
\vskip -1.1cm
}
\hskip -4.8cm
\vbox{
$\displaystyle{=~~\Tr(\Fge^a \Fge^b \Fge^b \Fge^a)}$
}
\hskip -9.15cm
\vbox{
$\displaystyle{=~~\frac{(\NC^2-1)^2}{4\NC}}$
}
}

\vskip 0.2cm

\hbox{
\hskip 0.5cm
\vbox{
\epsfxsize=3.2cm \epsffile{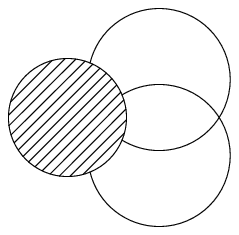}
\vskip -1.1cm
}
\hskip -4.8cm
\vbox{
$\displaystyle{=~~\Tr(\Fge^a \Fge^b \Fge^a \Fge^b)}$
}
\hskip -9.2cm
\vbox{
$\displaystyle{=~ - \frac{\NC^2-1}{4\NC}}$
}
}

\vskip 0.3cm

\hbox{
\hskip 0cm
\vbox{
\epsfxsize=4cm \epsffile{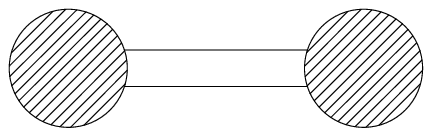}
\vskip -0.7cm
}
\hskip -4.7cm
\vbox{
$\displaystyle{=~~\Tr(\Fge^a \Fge^b)\Tr(\Fge^a \Fge^b)}$
}
\hskip -9.8cm
\vbox{
$\displaystyle{=~~\frac{\NC^2-1}{4}}$
}
}

\hbox{
\hskip 0.5cm
\vbox{
\epsfxsize=3.2cm \epsffile{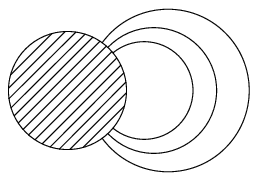}
\vskip -1.1cm
}
\hskip -4.4cm
\vbox{
$\displaystyle{=~~\Tr(\Fge^a \Fge^b \Fge^c \Fge^c \Fge^b \Fge^a)}$
}
\hskip -9.55cm
\vbox{
$\displaystyle{=~~\frac{(\NC^2-1)^3}{8\NC^2}}$
}
}

\hbox{
\hskip 0.5cm
\vbox{
\epsfxsize=3.2cm \epsffile{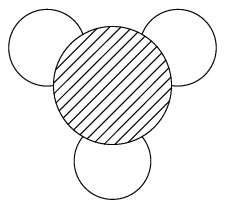}
\vskip -1.1cm
}
\hskip -4.4cm
\vbox{
$\displaystyle{=~~\Tr(\Fge^a \Fge^a \Fge^b \Fge^b \Fge^c \Fge^c)}$
}
\hskip -9.55cm
\vbox{
$\displaystyle{=~~\frac{(\NC^2-1)^3}{8\NC^2}}$
}
}

\vskip 0.6cm

\hbox{
\hskip 0.2cm
\vbox{
\epsfxsize=3.2cm \epsffile{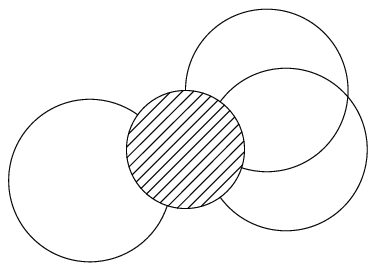}
\vskip -1.1cm
}
\hskip -4.1cm
\vbox{
$\displaystyle{=~~\Tr(\Fge^a \Fge^b \Fge^a \Fge^b \Fge^c \Fge^c)}$
}
\hskip -9.35cm
\vbox{
$\displaystyle{=~ - \frac{(\NC^2-1)^2}{8\NC^2}}$
}
}

\vskip 0.5cm

\hbox{
\hskip 0.6cm
\vbox{
\epsfxsize=3.2cm \epsffile{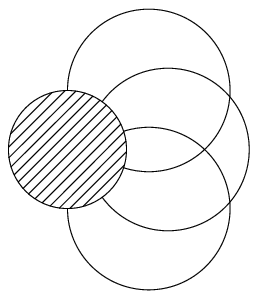}
\vskip -1.1cm
}
\hskip -4.4cm
\vbox{
$\displaystyle{=~~\Tr(\Fge^a \Fge^b \Fge^c \Fge^a \Fge^b \Fge^c)}$
}
\hskip -9.9cm
\vbox{
$\displaystyle{=~~\frac{\NC^2-1}{8\NC^2}}$
}
}

\vskip 0.7cm

\hbox{
\hskip 0.2cm
\vbox{
\epsfxsize=4cm \epsffile{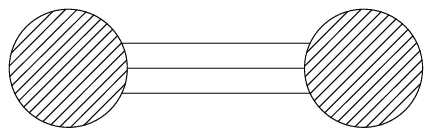}
\vskip -0.7cm
}
\hskip -4.4cm
\vbox{
$\displaystyle{=~~\Tr(\Fge^a \Fge^b \Fge^c)\Tr(\Fge^c \Fge^b \Fge^a)}$
}
\hskip -9.3cm
\vbox{
$\displaystyle{=~~\frac{(\NC^2-1)(\NC^2-2)}{8\NC}}$
}
}

\newpage

\hbox{
\hskip 0.2cm
\vbox{
\epsfxsize=4cm \epsffile{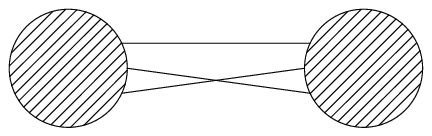}
\vskip -0.7cm
}
\hskip -4.4cm
\vbox{
$\displaystyle{=~~\Tr(\Fge^a \Fge^b \Fge^c)\Tr(\Fge^a \Fge^b \Fge^c)}$
}
\hskip -10.05cm
\vbox{
$\displaystyle{=~ - \frac{\NC^2-1}{4\NC}}$
}
}

\vskip 0.2cm

\hbox{
\hskip 0.2cm
\vbox{
\epsfxsize=4cm \epsffile{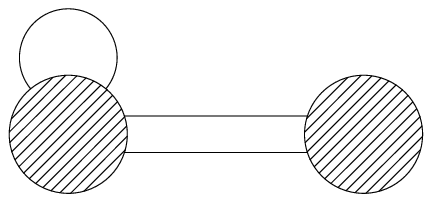}
\vskip -0.7cm
}
\hskip -4.4cm
\vbox{
$\displaystyle{=~~\Tr(\Fge^a \Fge^a \Fge^b \Fge^c)\Tr(\Fge^b \Fge^c)}$
}
\hskip -9.8cm
\vbox{
$\displaystyle{=~~\frac{(\NC^2-1)^2}{8\NC}}$
}
}

\vskip 0.1cm

\hbox{
\hskip 0.2cm
\vbox{
\epsfxsize=4cm \epsffile{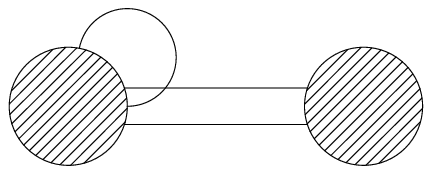}
\vskip -0.7cm
}
\hskip -4.4cm
\vbox{
$\displaystyle{=~~\Tr(\Fge^a \Fge^b \Fge^a \Fge^c)\Tr(\Fge^b \Fge^c)}$
}
\hskip -9.8cm
\vbox{
$\displaystyle{=~ - \frac{\NC^2-1}{8\NC}}$
}
}

\vskip 0.3cm

\hbox{
\hskip 0.6cm
\vbox{
\epsfxsize=3.2cm \epsffile{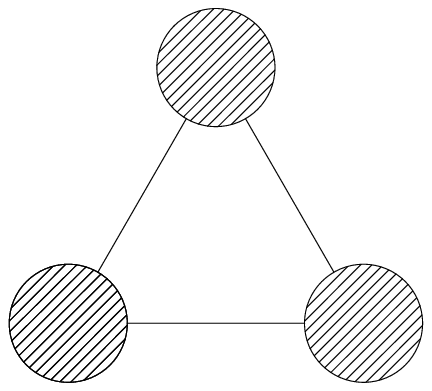}
\vskip -1.3cm
}
\hskip -4.1cm
\vbox{
$\displaystyle{=~~\Tr(\Fge^a \Fge^b)\Tr(\Fge^b \Fge^c)\Tr(\Fge^c \Fge^a)}$
}
\hskip -9.9cm
\vbox{
$\displaystyle{=~~\frac{\NC^2-1}{8}}$
}
}

\end{center}

\end {document}